\documentclass[prl,amssymb,amsmath,twocolumn,aps,showpacs]{revtex4}
\usepackage{amsmath} 
\usepackage{xr}
\usepackage{graphicx}
\usepackage{subfigure}
\newcommand{\dq}{\dot{q}}
\newcommand{\dw}{\dot{w}}
\usepackage[dvips]{epsfig}

\begin{document}

\title{Stochastic Efficiency for Effusion as a Thermal Engine}
\author{K. Proesmans}
\email{karel.proesmans@uhasselt.be}
\author{B. Cleuren}
\author{C. Van den Broeck}
\affiliation{Hasselt University, B-3590 Diepenbeek, Belgium}
\date{\today}

\begin{abstract}
The stochastic efficiency of  effusion as a thermal engine is investigated within the framework of stochastic thermodynamics. Explicit results are obtained for the probability distribution of the efficiency both at finite times and in the asymptotic regime of large deviations. The universal features, derived in Verley et al., Nature Communications 5, 4721 (2014), are reproduced. The effusion engine is a good candidate for both the numerical and experimental verification of these predictions.
\end{abstract}

\maketitle
\begin{figure}
\subfigure[]{\includegraphics[width=5cm]{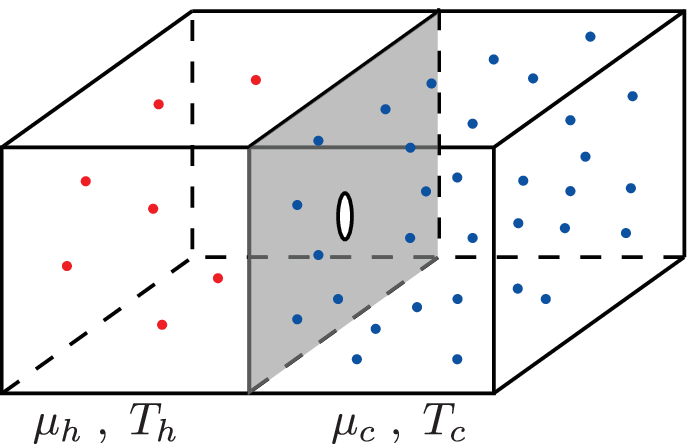}\label{fig.1a}}\\
\subfigure[]{\includegraphics[width=8cm]{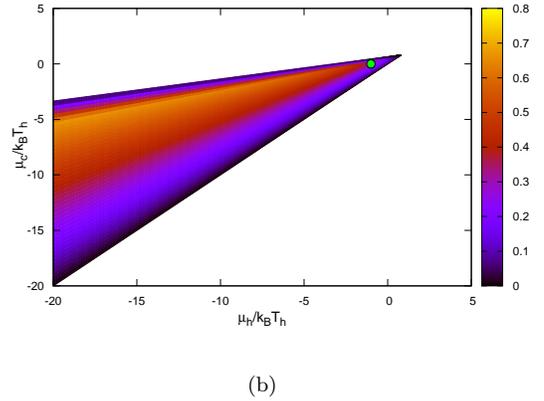}\label{fig mu}}
\caption{(Colour online) a) Effusion as a thermal engine b) Efficiency  $\bar{\eta}$ for an ideal (mono-atomic) gas, plotted in colour code as a function of $\mu_h/k_BT_h$ and $\mu_c/k_BT_h$ for $\eta_c=0.8$. The chemical potential is found from the Sackur-Tetrode formula, $\mu=k_BT\, ln\left(\rho\Lambda^{3}\right)$, where $\Lambda=h/\sqrt{2\pi m k_BT}$ is the thermal de Broglie wavelength, $k_B$ is  Boltzmann's constant, $T$  the temperature, $\rho$  the density and $m$ the mass per particle. The engine regime  ($T_h>T_c$) is determined by $\mu_h<\mu_c<(1-\eta_c)\mu_h-2 k_B T_c \ln(1-\eta_c)$.}
\label{fig1}
\end{figure}

Effusion is the escape of particles through a narrow aperture. The phenomenon has been used for many applications such as to enrich uranium, to coat light bulbs and as a cooling device. Effusion between two compartments, cf. Fig. \ref{fig.1a}, can also operate as a thermal engine, namely when there is a net flow of particles $n \geq 0$ from the hot compartment, temperature $T_h$, with low chemical potential $\mu_h$, to the cold compartment, temperature $T_c$, with high chemical potential $\mu_c$.  The work produced is $w=n \Delta \mu$ ($\Delta \mu=\mu_c-\mu_h$). 
If we denote by $q$ the net heat leaving  the hot reservoir, the resulting thermodynamic efficiency reads: 
\begin{equation}
\eta=w/q.
\end{equation}
When operating for long times $t$, the quantities  $n/t$ and $q/t$ converge to the average particle and heat flux. Concomitantly, the efficiency $\eta$ converges to its most probable value $\bar{\eta}$, which corresponds to the standard "macroscopic" efficiency. It is reproduced in Fig. \ref{fig mu} for the case of an ideal gas. As required by the second law  of thermodynamics, this long-time efficiency is always below the Carnot efficiency $\eta_c=1-T_c/T_h$.
\\
When operating for a finite time, the quantities $w$ and $q$, and hence also the {\it stochastic} efficiency $\eta$, are different from one run to another \cite{preCleuren}. One may wonder about thermodynamic implications for the probability distribution $P(\eta)$ of $\eta$. In a recent paper \cite{nc} (see also \cite{berkeley,verley2014universal}), the following remarkable result was derived:
in a non-macroscopic machine running for long but finite times, the reversible efficiency is least probable (in the sense of large deviations \cite{touchette2009large}). This property is based on the generalisation of the second law of thermodynamics for small systems, the so-called fluctuation theorem, and parallels the derivation of Carnot or reversible efficiency of macroscopic machines. The simplest illustration is the work to work transformation by a Brownian particle subjected to competing external forces \cite{nc}. The reversible efficiency is here $100\%$. Both work components are Gaussian and the explicit analytic result is available for the probability distribution of the stochastic efficiency \cite{polettini2014finite} and its large deviation function \cite{nc}. In this letter, we calculate the stochastic efficiency for the effusion engine, which is, as we argue below, the simplest possible construction of a thermal machine. The analytic expression for the large deviation properties of work and heat can be obtained, as well as that for the efficiency apart from a final step involving a parametric elimination. Furthermore, the finite time regime with the approach to the large deviation limit and the deviations from the Gaussian regime can be studied in detail. All the results are of course in agreement with the universal predictions from \cite{nc}.

The simplicity of the effusion engine stems from the fact that there is no auxiliary engine part that transfers the energy. Furthermore, the crossing of the particles is described by basic Poisson statistics. To illustrate the calculations and to incidentally show  how the Carnot efficiency can be recovered, we first consider effusion operating with a filter, such that only particles with energies in a small window $ ]E-\Delta E/2,E+\Delta E/2[$ are allowed to cross. The number $n_h$ of particles leaving the hot reservoir during a time $t$ is described by a Poisson distribution:
\begin{equation}
P(n_h,t)=\frac{\bar{n}_h^{n_h}}{n_h!} e^{-\bar{n}_h}\;\;,\;\;\bar{n}_h=k t,
\end{equation}
with the crossing rate $k$ prescribed by kinetic theory:
\begin{equation}
k=\frac{1}{t_0}\frac{E \Delta E}{\left(k_B T_h\right)^2}e^{-E/k_B T_h}.
\end{equation}
Here we introduced the average escape time $t_0$ for a particle in the absence of an energy filter:
\begin{equation}
t_0=\frac{\sqrt{2\pi m}}{\sigma\rho_h\sqrt{k_BT_h}},
\end{equation}
and $\sigma$ is the surface area of the effusion hole.
A similar result holds for the crossing of particles coming from the cold reservoir by formally replacing the subscript $h$ by $c$ and $k$ by $l$.
Since the crossings are independent from each other, we find by convolution that the probability to have a net transfer of $n=n_h-n_c$ particles, is given by:
\begin{equation}
P(n,t)=e^{-t(k+l)}\left(\frac{k}{l}\right)^{\frac{n}{2}}I_{n}\left(2t\sqrt{kl}\right).\label{pn}
\end{equation}
This is nothing but the probability distribution for a biased continuous-time random walk with stepping rates $k$ and $l$.
The so-called large deviation function $\varphi(\dot{n})$ describes the asymptotic probability for observing an empirical particle flux  $\dot{n}=n/t$:
\begin{equation}
 P(n=\dot{n}t,t)\sim e^{-t\varphi(\dot{n})}\;\;\; \mbox{or}\;\;\;\varphi(\dot{n})=-\underset{t \rightarrow \infty}{\lim}\frac{\ln P(n=\dot{n}t,t)}{t}.
\end{equation}
It is found from Eq. (\ref{pn}) by Stirling's formula:
\begin{equation}
\varphi(\dot{n})=k+l-\sqrt{4kl+\dot{n}^2}-\dot{n}\ln\left[\frac{\sqrt{4kl+\dot{n}^2}-\dot{n}}{2l}\right].\label{NLDF}
\end{equation}
Note that in the presence of a single energy filter, the net energy transfer $u$ and net particle transfer $n$ become "strongly coupled": $u=E n$. Also work  $w=n \Delta\mu$ and heat $q=u-n \mu_h=n (E-\mu_h)$ are proportional to each other. Hence, even while both $w$ and $q$ fluctuate through their dependence on $n$, the efficiency, $\eta= w/q$, remains constant.  Since in this case $\eta=\bar{\eta}$, the second law  requires  $\eta \leq \eta_c$. It is revealing to show how Carnot efficiency can be reached in this case \cite{prlLinke,VdB}. The crucial condition is that one operates under equilibrium conditions, i.e., the Maxwell-Boltzmann density of particles of energy $E$ has to be the same in both compartments: $\rho_h \exp(-\beta_h E)/T_h^{3/2}=\rho_c \exp(-\beta_c E)/T_c^{3/2}$. This indeed yields for an ideal gas ($\mu\sim T\ln(\rho/T^{3/2})$ apart from an additive constant):
$\eta=w/q=(\mu_c-\mu_h)/(E-\mu_h)=\eta_c$ (see Supplemental Material). Note that equilibrium is reached even though density, chemical potential and temperature need not be the same in both compartments.

To obtain a nontrivial result for the probability distribution $P_t(\eta)$ of the efficiency, we consider next effusion through two separate windows, selective for energies $E_1$ and $E_2$, respectively. 
The discrete amounts of work and heat, produced upon a net transfer (from hot to cold) of $i$ particles through window $1$ and $n-i$ particles through window $2$, are obviously given by:
\begin{equation}
w=n\Delta \mu\;,\;\;q=\delta q_1i+\delta q_2(n-i),
\end{equation}
with $\delta q_1$ and $\delta q_2$ the transported heat per particle leaving the hot reservoir via filter $1$ and $2$, respectively :
\begin{equation}
\delta q_1= E_1-\mu_h\;\;,\;\;\delta q_2= E_2-\mu_h.
\end{equation}
Since the transport through both filters is statistically independent, the  joint probability $P_t(w,q)$ for work and heat is given by the following convolution:
\begin{eqnarray}\label{ftwq}
P_t(w,q)&=&\sum_{i_1,i_2} P_1(i_1,t)P_2(i_2,t)\delta^{Kr}_{q,\delta q_1 i_1+\delta q_2 i_2}\delta^{Kr}_{w,\Delta\mu (i_1+i_2)}\nonumber\\
&=&P_1\left(\frac{q-\delta q_2 n}{\delta q_1-\delta q_2},t\right)P_2\left(\frac{\delta q_1 n-q}{\delta q_1-\delta q_2},t\right),
\end{eqnarray}
$P_j(n,t)$  being the probability for a net transfer of $n$ particles through the $j$-th filter after a time $t$, $j=1,2$. It is given by Eq. (\ref{pn}) with $E$ and $\Delta E$ replaced by the corresponding values for filter $j$ (we used the same $\sigma$ for both filters). Considering the large time limit, the heat and work fluxes $\dq=q/t$ and $\dw=w/t$ are characterised by the following large deviation function:
\begin{eqnarray}
I(\dot{w},\dot{q})&=&-\underset{t\rightarrow\infty}{\lim}\frac{\ln P_t (w=\dw t,q=\dq t)}{t}\nonumber\\&=&\varphi_1\left(\frac{\dot{q}-\delta q_2 \dot{w}/\Delta\mu}{\delta q_1-\delta q_2}\right)+\varphi_2\left(\frac{\dot{q}-\delta q_1 \dot{w}/\Delta\mu}{\delta q_2-\delta q_1}\right),\nonumber\\\label{I2F}
\end{eqnarray}
where $\varphi_j$ is the large deviation function of the net particle flux through the $j$-th filter, $j=1,2$, given by \eqref{NLDF} with the appropriate values of the rates $k$ and $l$.

Turning to the efficiency $\eta=w/q$, we note that its probability distribution:
\begin{eqnarray}
P_t(\eta)&=&\int_{-\infty}^{\infty}\int_{-\infty}^{\infty}P_t(w,q)\delta(\eta-w/q)dwdq\nonumber\\&=&\int_{-\infty}^{\infty}P_t(w,w/\eta)\left|\frac{w}{\eta^2}\right|dw,\label{Peta}
\end{eqnarray}
can be obtained for any finite time by combination with Eq.(\ref{ftwq}), cf. Fig. \ref{figPMF}.
For the corresponding large deviation function  $J(\eta)$, one finds:
\begin{figure}
\subfigure[]{\includegraphics[width=8cm]{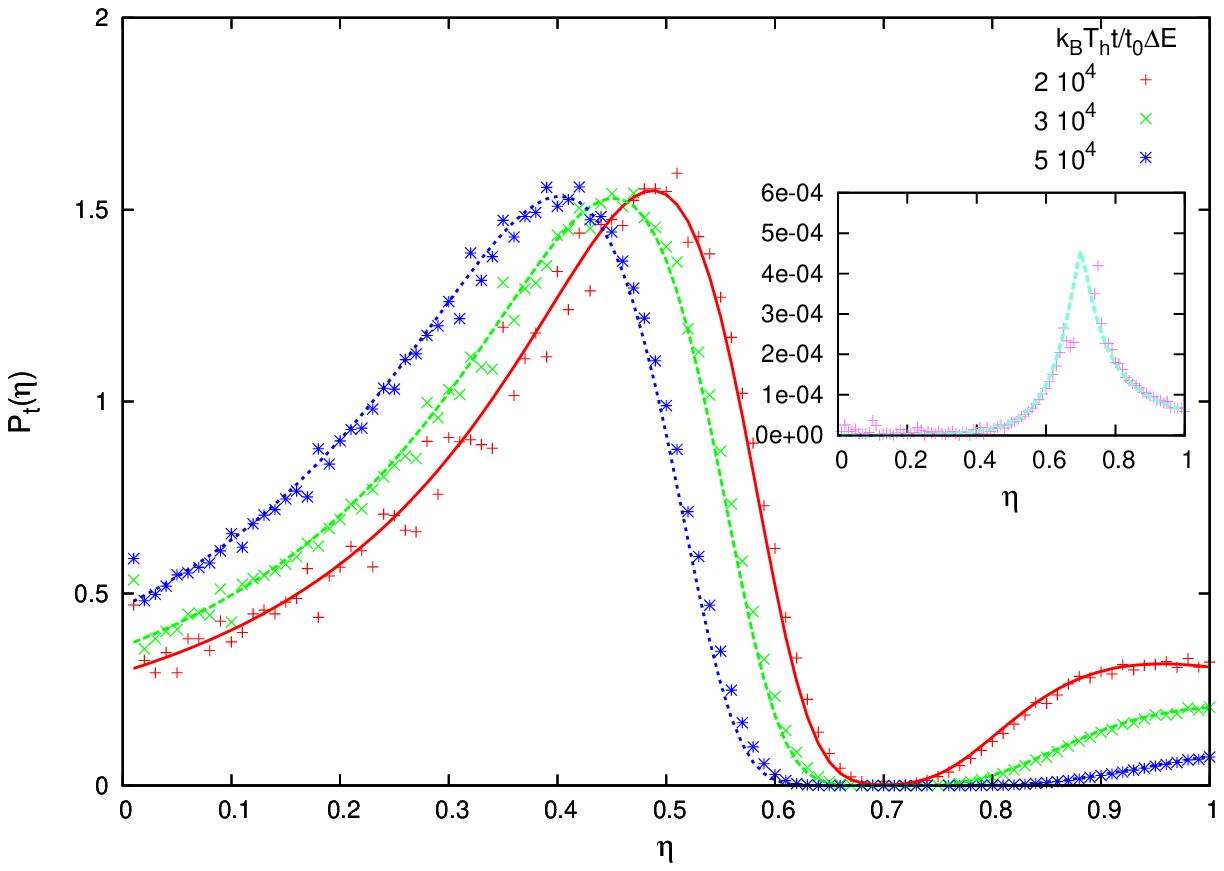}\label{figPMF}}\\
\subfigure[]{\includegraphics[width=8cm]{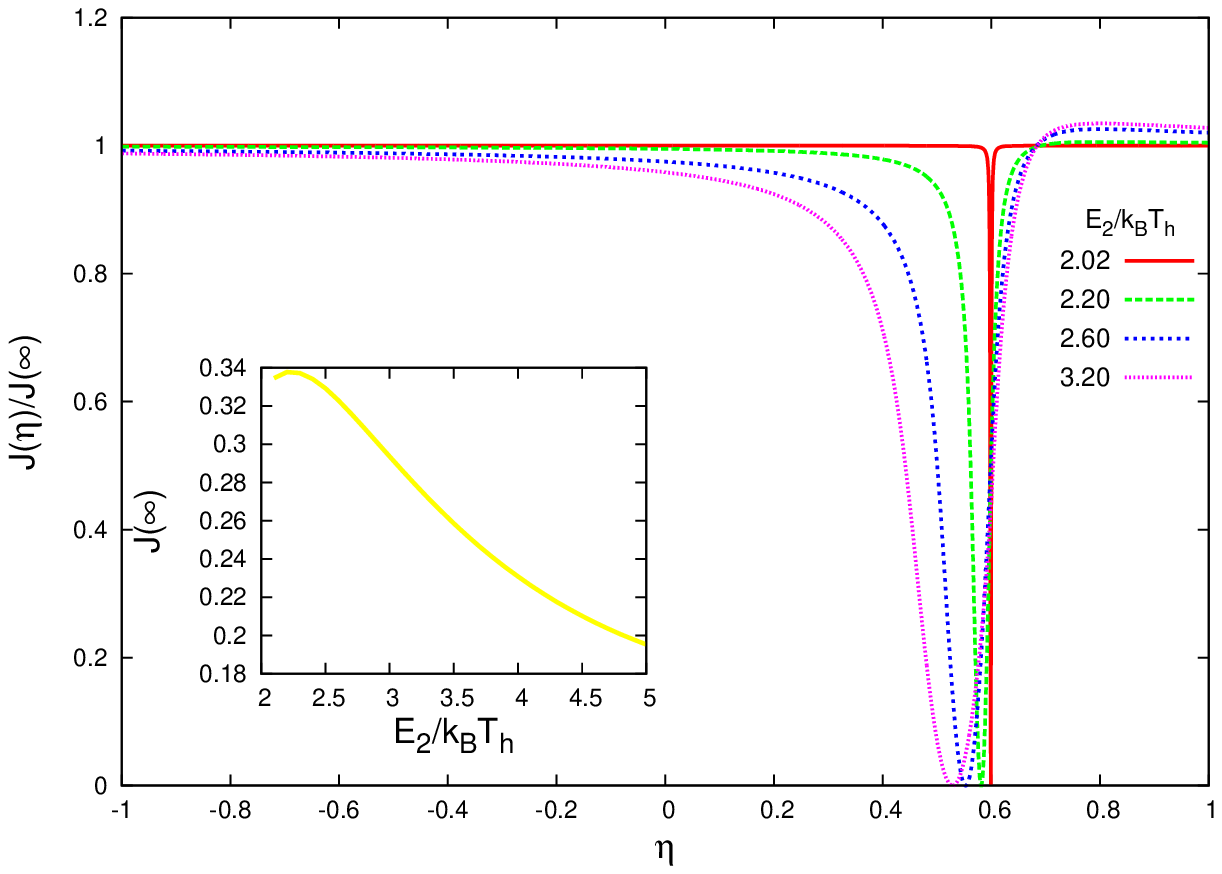}\label{figsc}}
\caption{(Colour online) Results for effusion with filters: a) $P_t(\eta)$ for $t/t_0 \,\Delta E /k_B T_h$ = $2\,10^4$, $3\,10^4$ and $5\,10^4$, after $8\,10^7$ runs. Full lines: $P_t(\eta)$ from Eq. \eqref{Peta}. Parameter values: $\eta_c=0.7$, $\mu_h=-k_B T_h$, $\mu_c=0.05\, k_B T_h$, $E_1 =0.5\, k_B T_h$, $E_2=10\,k_B T_h$. Inset: $J(\eta)$ exact result (dashed line) compared to extrapolation from finite time (crosses) (see Supplemental Material).  b) $J(\eta)/J(\infty)$ for different values of $E_2$. Parameter values: $\eta_c=0.8$, $E_1=2\,k_BT_h$, $\mu_h=-k_B T_h$, $\mu_c=0.8\,k_B T_h$. Note that the system becomes "strongly coupled" in the limit $E_2\rightarrow E_1$  Inset: $J(\infty)$ as a function of $E_2$. }\label{figMF}
\end{figure}
\begin{eqnarray}
J(\eta)&=&-\underset{t\rightarrow\infty}{\lim}\frac{\ln\left(P_t(\eta)\right)}{t}= \underset{\dot{q}}{\text{min}}\;I(\eta \dot{q},\dot{q})\nonumber \\
&=& \underset{\dot{q}}{\text{min}}\;\left\{\varphi_1\left(\frac{\gamma_1\left(\eta\right)}{\Delta \mu} \dot{q}\right)+\varphi_2\left(\frac{\gamma_2\left(\eta\right)}{\Delta \mu} \dot{q}\right)\right\},\label{Ieta}
\end{eqnarray}
with 
\begin{equation}
\gamma_1(\eta)=\frac{\Delta \mu-\delta q_2 \eta}{\delta q_1-\delta q_2} \;\;\;\; ; \;\;\;\; \gamma_2(\eta)=\frac{\Delta \mu-\delta q_1 \eta}{\delta q_2-\delta q_1}.
\end{equation}
We have  used the contraction principle \cite{touchette2009large}, expressing the fact that a given value of $\eta$ is realised by the most likely values of $\dot{w}$ and $\dot{q}$ for which $\dot{w}/\dot{q}=\eta$. 
The above minimization involves a transcendental equation (see Supplemental Material), requiring a numerical solution. The resulting large deviation function $J(\eta)$ has the  familiar shape \cite{nc}, with a zero in $\bar{\eta}$, a maximum at Carnot efficiency, and equal asymptotes for $\eta \rightarrow \pm \infty$, cf.  Fig. \ref{figsc}.

\begin{figure}
\subfigure[]{\includegraphics[width=8cm]{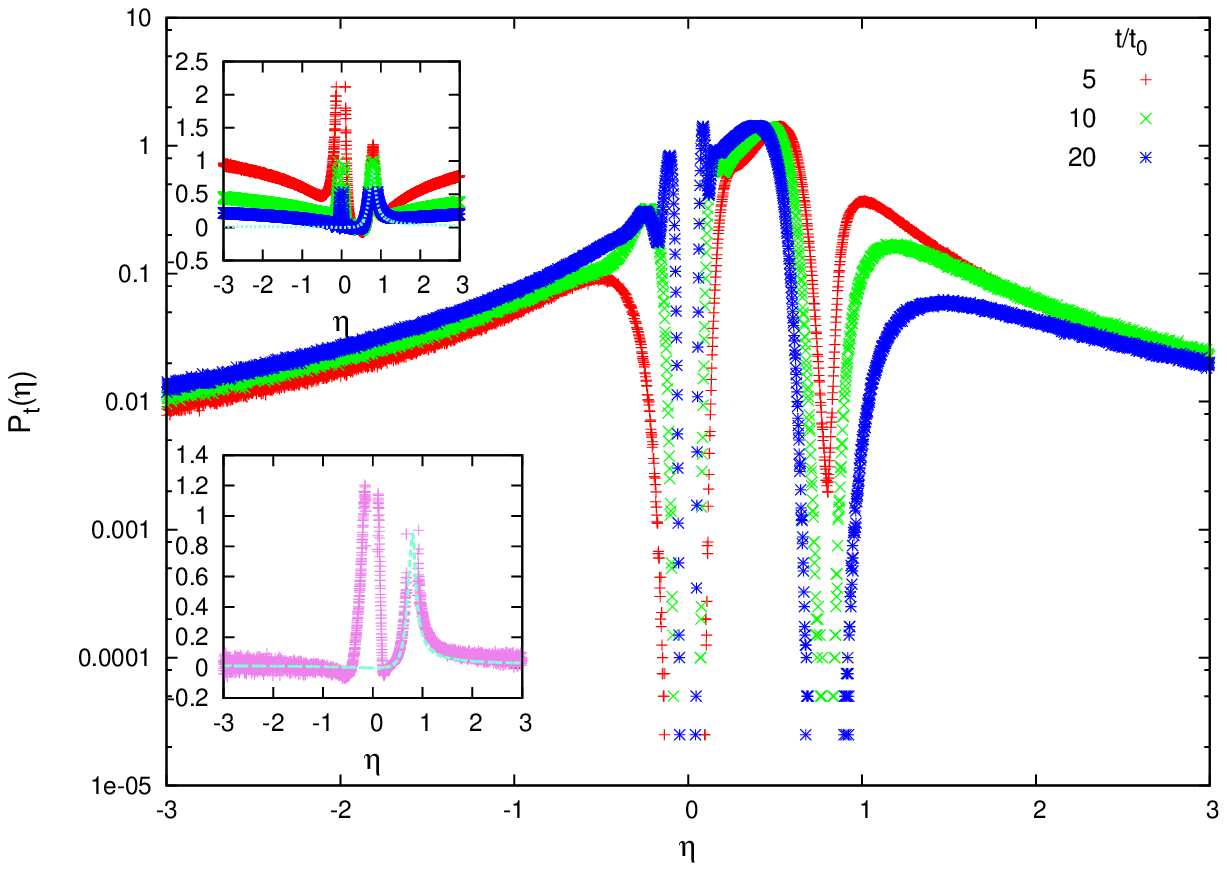}\label{figPJZF}}\\
\hfill
\subfigure[]{\includegraphics[width=8cm]{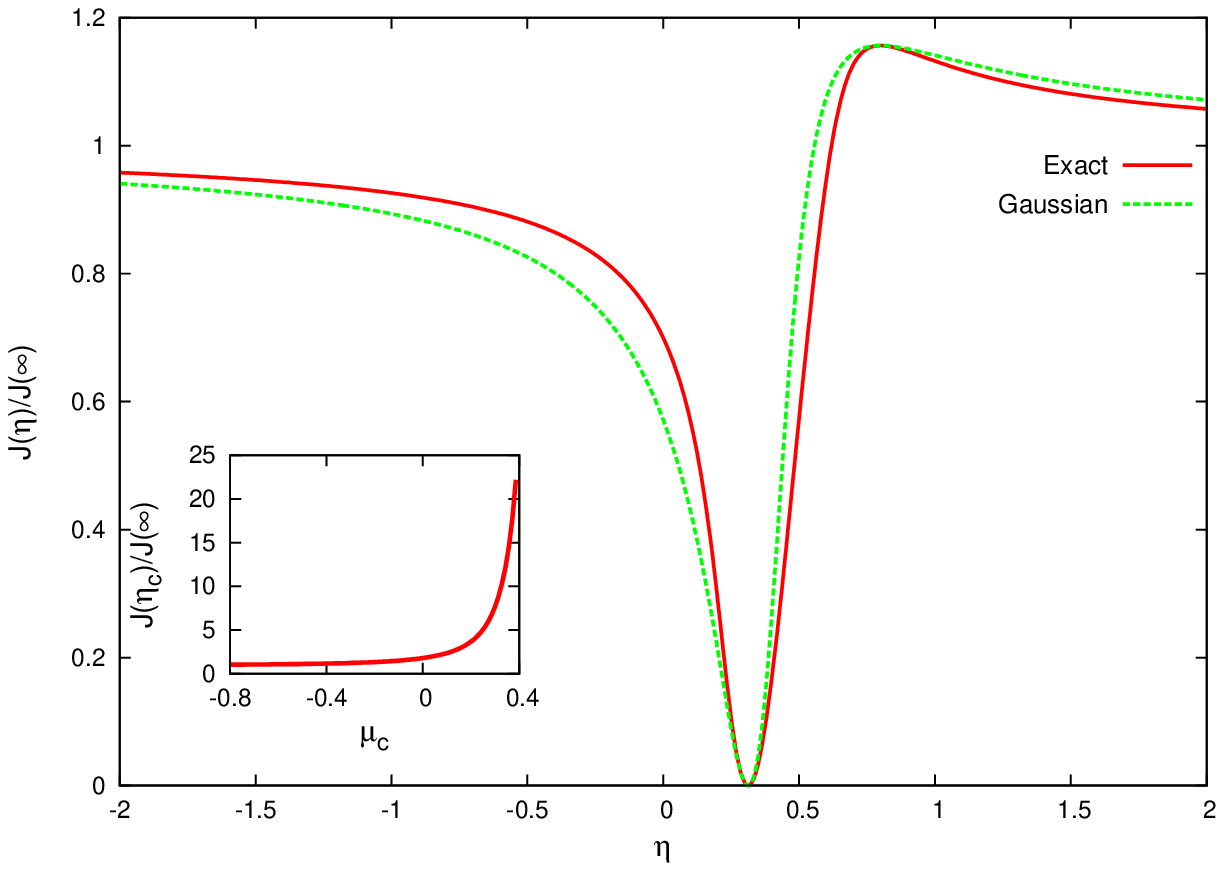}\label{figNOF}}
\caption{(Colour online) Results for plain effusion (no filters): a)   $P_t(\eta)$ for $t/t_0 = 5, 10, 20$, after $2\, 10^7$ runs. Parameter values: $\eta_c=0.7$, $\mu_h=-5\,k_B T_h$, $\mu_c=\mu_h (1-\eta_c) -2\,k_B T_c \ln(1-\eta_c)$. Upper inset: approach to the large deviation limit, $-\ln(P_t(\eta))/t$ for $t/t_0 = 5, 10, 20$. Lower inset: $J(\eta)$ exact result (dashed line) compared to extrapolation from finite time (crosses). b) $J(\eta)/J(\infty)$ for $\eta_c=0.8$, $\mu_h=-k_B T_h$ and $\mu_c=0$ (dot in Fig. \ref{fig mu}). Inset: $J(\eta)/J(\infty)$ as a function of $\mu_c$ in the engine regime.}
\label{figZF}
\end{figure}

One can repeat the above calculation for effusion without energy filters (see Supplemental Material). Results are shown in Fig. \ref{figZF}.  Note in Fig. \ref{figPJZF} the "fine structure" appearing for small times around $\eta=0$, due to the fact that very few particles will cross. The observed minimum of $P_t(\eta)$ around zero is for example due to the very unlikely single particle crossing with high energy, the latter being required to obtain a small value of $\eta$.

\begin{figure}
\includegraphics[width=8cm]{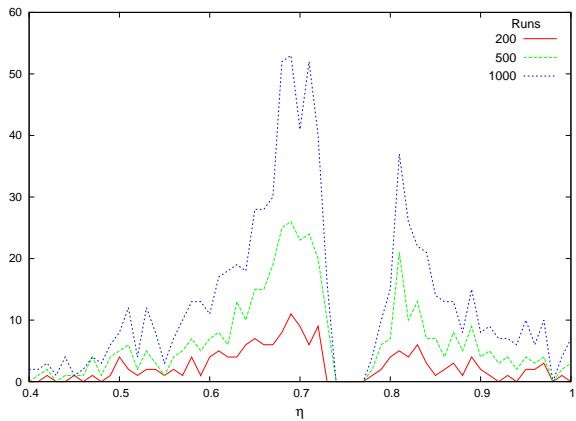}
\caption{Number of events in function of $\eta$ (sample width $d\eta=0.01$) for 200, 500 and 1000 runs. Parameter values: $E_1=k_B T_h, E_2=10\, k_B T_h, \text{P}_c=6.3\, 10^4Pa,  T_c=100\,K, \text{P}_h=1.0\, 10^5Pa, T_h=400\,K, \sigma=100\,nm^2$ and $\Delta E=k_B T_h/1000$. The running time is $t=7.6 \,10^{-5} sec$ (value for Helium).}
\label{FigFS}
\end{figure}

We turn to a discussion of the salient features and implications of our analysis.
 The first conclusion is that  all the predictions of the general theory \cite{nc} are verified, and in particular the generic properties of the large deviation function $J(\eta)$. Second, both the Gaussian regime and the strongly non-Gaussian regime can be easily observed, as well as other special limits such as the "strong coupling" limit or the limit to "reversibility". 
 Third, the minimum of the probability or the maximum of the large deviation function at Carnot efficiency can be made very pronounced, rendering the most striking feature, the minimum at Carnot efficiency, easy to observe. Fourth, an accurate estimation of the large deviation function $J(\eta)$ is possible by extrapolating the results from finite time, as is shown in the insets of Figs. \ref{figPMF} and \ref{figPJZF}.
 Fifth, even for relatively short times (less than $10^{-4}$ sec) and with rather small statistics ($200$ samples), the minimum of the the probability $P_t(\eta_c)$ can clearly be identified, as can be seen in Fig. \ref{FigFS}.
 Sixth, the probability distribution for the efficiency of effusion displays long tails, similar to the one observed in the Gaussian scenario \cite{polettini2014finite}. This appears to be a generic result. Indeed it follows from \eqref{Peta} that:
\begin{equation}P_t (\eta)\approx\frac{\int P_t (w,0)\left|w\right|dw}{\eta^2},\end{equation}
provided $P_t (w,q)$ has a smooth behavior around $q=0$ and the integral converges.
Last but not least, we discuss the experimental setting for an effusion engine displaying the salient features of stochastic efficiency. To obtain a sharp minimum of $P_t(\eta)$ at Carnot efficiency, a large value of $J(\eta_c)/J(\infty)$ is required. One way to do so in the two filter effusion engine is by considering $\mu_c\approx \mu_hT_c/T_h +E_1\eta_c$ and $E_2\gg E_1$. This region is described by $\text{P}_{h}/\text{P}_{c}=\left(T_{h}/T_c\right)^{5/2}e^{-\eta_c/(1-\eta_c)}$ (see Supplemental Material).
With this condition in mind, we have performed simulations for the stochastic efficiency of an effusion engine operating with  Helium gas.  In Fig. \ref{FigFS}, we represent $P_t(\eta)d\eta$ (window $d\eta=.01$) obtained by  $200$, $500$ and $1000$ samplings. The minimum in the vicinity of Carnot efficiency $\eta_C=0.75$ is clearly visible, even at these very short times.

In conclusion,  effusion as a thermal engine displays all the key features of  stochastic efficiency. The model combines conceptual simplicity with analytic tractability  and experimental relevance. It is generic for the case in which transitions  are  ruled by Poisson statistics, such as in Kramers' escape over a potential barrier, or for transitions ruled by a chemical reaction. 

%

\end{document}


\title{Stochastic Efficiency for Effusion as a Thermal Engine\\Supplemental Notes}
\author{K. Proesmans}
\email{karel.proesmans@uhasselt.be}
\author{B. Cleuren}
\author{C. Van den Broeck}
\affiliation{Hasselt University, B-3590 Diepenbeek, Belgium}
\date{\today}
\maketitle

\section{Reaching Carnot efficiency}
Consider an ideal gas effusion engine  operating with a filter at an energy E such that the density of particles with this energy is the same in both compartments:
\begin{equation}\frac{n_h}{T_h ^{3/2}}e^{-\frac{E}{k_BT_h}}=\frac{n_c}{T_c ^{3/2}}e^{-\frac{E}{k_BT_c}}.\end{equation}
Using the Sackur Tetrode formula for the chemical potential, we find:
\begin{eqnarray}E&=&\frac{k_BT_hT_c}{T_h-T_c}\ln\left(\frac{n_c T_h^{3/2} }{n_h T_c ^{3/2}}\right)=\frac{T_h\mu_c-T_c \mu_h}{T_h-T_c},\end{eqnarray}
so that
\begin{eqnarray}\bar{\eta}&=&\frac{\mu_c -\mu_h}{E-\mu_h}=1-\frac{T_c}{T_h}\end{eqnarray}
For free effusion, we have:
\begin{eqnarray}\bar{\eta}&=&\frac{(\mu_c-\mu_h) \left((k_BT_h)^2 e^{\frac{\mu_h}{k_BT_h}}-(k_BT_c)^2 e^{\frac{\mu_c}{k_BT_c}}\right)}{(k_BT_h)^2 e^{\frac{\mu_h}{k_BT_h}} (2 k_BT_h-\mu_h)-(k_BT_c)^2 e^{\frac{\mu_c}{k_BT_c}} \left(2 k_BT_c-\mu_h \right)}\nonumber\\&\leq & \frac{(\mu_c-\mu_h) \left((k_BT_h)^2 e^{\frac{\mu_h}{k_BT_h}}-(k_BT_c)^2 e^{\frac{\mu_c}{k_BT_c}}\right)}{(2k_BT_h-\mu_h)\left((k_BT_h)^2 e^{\frac{\mu_h}{k_BT_h}}-(k_BT_c)^2 e^{\frac{\mu_c}{k_BT_c}}\right)}\nonumber\\&=&\frac{\mu_h-\mu_c}{\mu_h-2k_B T_h}\nonumber\\&\leq &\eta_c,\end{eqnarray}
where we made use of the fact that one operates in the heat engine regime. Carnot efficiency is only reached in the limit:
\begin{equation}\mu_h\rightarrow-\infty\;\; ,\;\;\mu_c\rightarrow (1-\eta_c)\mu_h.\end{equation}

\section{The large deviation function by extrapolation}
To  estimate the LDF $J(\eta)$ from finite time measurements of $P_t(\eta)$, we  use the following form:
\begin{equation}P_t(\eta)=A(\eta)t^{B(\eta)}\exp(-tJ(\eta)).\end{equation}
To fit the values of $A(\eta)$, $B(\eta)$ and $J(\eta)$, it is sufficient to have a measurement of $P_t(\eta)$ at $3$ different times. More precisely, from the values:
\begin{equation}\tau_i(\eta)=-\frac{1}{t_i}\ln(P_{t_i}(\eta)),\;\;\;i=1,2,3,\end{equation}
we get:
\begin{equation}\left[\begin{array}{c}\tau_1\\\tau_2\\\tau_3\end{array}\right]= \left[ \begin {array}{ccc} 1&{t_{{1}}}^{-1}&{\frac {\ln  \left( t_{{1
}} \right) }{t_{{1}}}}\\ \noalign{\medskip}1&{t_{{2}}}^{-1}&{\frac {
\ln  \left( t_{{2}} \right) }{t_{{2}}}}\\ \noalign{\medskip}1&{t_{{3}}
}^{-1}&{\frac {\ln  \left( t_{{3}} \right) }{t_{{3}}}}\end {array}
 \right] \left[\begin{array}{c}J(\eta)\\A(\eta)\\B(\eta)\end{array}\right].\end{equation}
By inversion of the matrix, we find the required estimate of $J(\eta)$:
\begin{equation}J(\eta)={\frac {t_{{1}}\tau_{{1}}\ln  \left( t_{{2}} \right) -t_{{3}}\tau_{{3}
}\ln  \left( t_{{2}} \right) -t_{{1}}\tau_{{1}}\ln  \left( t_{{3}}
 \right) +t_{{2}}\tau_{{2}}\ln  \left( t_{{3}} \right) -t_{{2}}\tau_{{
2}}\ln  \left( t_{{1}} \right) +t_{{3}}\tau_{{3}}\ln  \left( t_{{1}}
 \right) }{\ln  \left( t_{{2}} \right) t_{{1}}-\ln  \left( t_{{2}}
 \right) t_{{3}}-\ln  \left( t_{{1}} \right) t_{{2}}+\ln  \left( t_{{1
}} \right) t_{{3}}-\ln  \left( t_{{3}} \right) t_{{1}}+\ln  \left( t_{
{3}} \right) t_{{2}}}}
\end{equation}

\section{$J(\eta)$ in the presence of two filters.}
$J(\eta)$ in the presence of two filters is given by the following expression:
\begin{eqnarray}
J(\eta)&=&\lim_{t\rightarrow\infty}\frac{\ln\left(P_t(\eta)\right)}{t}= \min_{\dot{q}}\;I(\eta \dot{q},\dot{q})\nonumber \\
&=& \min_{\dot{q}}\;\left\{\varphi_1\left(\frac{\gamma_1\left(\eta\right)}{\Delta \mu} \dot{q}\right)+\varphi_2\left(\frac{\gamma_2\left(\eta\right)}{\Delta \mu} \dot{q}\right)\right\},\label{Ieta2}
\end{eqnarray}
with 
\begin{equation}
\gamma_1(\eta)=\frac{\Delta \mu-\delta q_2 \eta}{\delta q_1-\delta q_2} \;\;\;\; ; \;\;\;\; \gamma_2(\eta)=\frac{\Delta \mu-\delta q_1 \eta}{\delta q_2-\delta q_1}.
\end{equation}
The minimization with respect to $\dot{q}$ is  done numerically by solving:
\begin{eqnarray}
\gamma_1(\eta) \ln\left(\frac{\sqrt{4k_1l_1+(\frac{\gamma_1(\eta)}{\Delta \mu} \dot{q})^2}-\frac{\gamma_1(\eta)}{\Delta \mu} \dot{q}}{2l_1}\right)+\gamma_2(\eta) \ln\left(\frac{\sqrt{4k_2l_2+(\frac{\gamma_2(\eta)}{\Delta \mu} \dot{q})^2}-\frac{\gamma_2(\eta)}{\Delta \mu} \dot{q}}{2l_2}\right)=0.\nonumber\\
\end{eqnarray}

\section{$J(\eta)$ without filters.}
In \cite{verley2014universal}, one derives the following result:
\begin{equation}
J(\eta)=-\min\limits_{\lambda}\varphi(\lambda,\lambda \eta),
\end{equation}
where $\varphi(\lambda,\omega)$ is the cumulant generating function of  work and heat. For effusion without filters, it is given by \cite{preCleuren}:
\begin{eqnarray}
\varphi(\lambda,\omega)=-\frac{\sigma \rho_h \sqrt{k_BT_h}}{\sqrt{2\pi m}}\left(1-\frac{e^{-\lambda \Delta \mu - \omega \mu_h}}{(1-k_BT_h \omega)^2} \right)-\frac{\sigma \rho_c \sqrt{k_BT_c}}{\sqrt{2\pi m}}\left(1-\frac{e^{\lambda \Delta \mu + \omega \mu_c}}{(1+k_BT_c \omega)^2} \right).
\end{eqnarray}
The results presented in Fig. \ref{figZF} are obtained by the performing the above minimization numerically.
\section{Maximizing $J(\eta_c)/J(\infty)$.}
We consider effusion in the presence of two filters with $E_2\gg E_1$ and $\mu_c\approx\mu_h (1-\eta_c)+E_1\eta_c$. This implies $k_1\gg k_2\gg k_1-l_1\gg l_2$. Hence one has approximately $l_1\approx k_1$, i.e., the net particle flux through the hole with energy $E_1$ is approximately zero, and $l_2\approx 0$. The large deviation functions of the net particle transport are then given by:
\begin{equation}\varphi_1 (\dot{n}) \approx 2k_1 -\sqrt{4k_1 ^2+\dot{n}^2}-\dot{n}\,\ln\left(\frac{\sqrt{4k_1^2+\dot{n}^2}-\dot{n}}{2k_1}\right), \end{equation}
\begin{equation}\varphi_2 (\dot{n}) \approx k_2 -\dot{n}-\dot{n}\,\ln\left(\frac{k_2}{\dot{n}}\right).\end{equation}
We thus find:
\begin{eqnarray}J(\infty)&=&\min_\lambda I(0,\lambda)=\min_\lambda\left( \varphi_1 (\lambda)+\varphi_2(-\frac{\delta q_1}{\delta q_2}\lambda)\right)\leq \varphi_1\left(-\frac{\delta q_2}{\delta q_1}k_2\right)+\varphi_2(k_2).\end{eqnarray}
Since:
\begin{equation}
\varphi_2(k_2)\approx 0,
\end{equation}
one finds, using a Taylor approximation around $\dot{n}=0$:
\begin{equation}J(\infty)\leq\varphi_1 \left(-\frac{\delta q_2}{\delta q_1}k_2\right)\approx\frac{1}{4 k_1}\left(\frac{\delta q_2 k_2}{\delta q_1}\right)^2.\end{equation}
Combined with:
\begin{equation}
J(\eta_c)=I(0,0)=k_1+l_1-2\sqrt{k_1 l_1}+k_2+l_2-2\sqrt{k_2 l_2}\approx k_2,
\end{equation}
one concludes:
\begin{equation}\frac{J(\eta_c)}{J(\infty)}\geq 4\left(\frac{\delta q_1}{\delta q_2}\right)^2\frac{k_1}{k_2},\end{equation}
which can become arbitrarily large for $E_2\gg E_1$.
\section{Effusion engine operating with Helium.}
From the mass of a Helium atom:
\begin{equation}m_{helium} =6.65\,10^{-27}\,\mathrm{kg}.\end{equation}
we find the corresponding thermal de Broglie wavelength:
\begin{equation}\Lambda_{helium}=\frac{8.87\,10^{-10}}{\sqrt{T}}\,\mathrm{K^{1/2}m}.\end{equation}
The relation between chemical potential, temperature and density thus becomes:
\begin{equation}\rho=\frac{N}{V}=1.50\,10^{27}\,T^{3/2}e^\frac{\mu}{k_BT}\mathrm{K^{-3/2}m^{-3}},\end{equation}
or in terms of the pressure:
\begin{equation}P=\rho k_BT=T^{5/2}\,e^{\mu/k_BT+9.94}\,\mathrm{K^{-5/2}Pa}\end{equation}
If we set:
\begin{equation}\mu_c=(1-\eta_c)\mu_h+E_1\eta_c,\end{equation}
and $E_1=kT_h$, we get:
\begin{equation}P_{h}=T_{h}^{5/2}e^{\frac{\mu_h}{k_BT_h}+9.94}\,\mathrm{K^{-5/2}Pa},\end{equation}
\begin{equation}P_{c}=T_{c}^{5/2}e^{\frac{\mu_h}{k_BT_h}+\frac{T_h}{T_c}\eta_c+9.94}\,\mathrm{K^{-5/2}Pa}.\end{equation}
Note that for the Sackur-Tetrode and the ideal gas law to be valid, we should have:
\begin{equation}e^{\frac{\mu}{k_BT}}\ll 1.\end{equation}
For:
\begin{eqnarray}P=10^5\,\mathrm{Pa}, \;\;\;\;T=400\,\mathrm{K},\end{eqnarray}
we have:
\begin{equation}\mu=-13.40 k_BT,\end{equation}
which indeed lies in the 'ideal gas' regime.
%